\documentclass[aps,prl,preprintnumbers,amsmath,amssymb,latexsym,array,enumerate,letter,twocolumn,superscriptaddress]{revtex4}
\usepackage{amssymb}
\usepackage{amsmath}
\usepackage{epsfig}
\usepackage{hyperref}
\usepackage{breakurl}
\usepackage{mathtools}
\usepackage{extpfeil}
\usepackage{comment}
\usepackage{multirow}
\usepackage{graphicx,float,color}
\usepackage{url}
\usepackage{ulem} 
\usepackage{newfile}
\makeatletter
\def\simgt{\mathrel{\lower2.5pt\vbox{\lineskip=0pt\baselineskip=0pt
           \hbox{$>$}\hbox{$\sim$}}}}
\def\simlt{\mathrel{\lower2.5pt\vbox{\lineskip=0pt\baselineskip=0pt
           \hbox{$<$}\hbox{$\sim$}}}}
\makeatother

\def\lsim{\mathrel{\rlap{\lower4pt\hbox{\hskip1pt$\sim$}}
     \raise1pt\hbox{$<$}}}         
\def\gsim{\mathrel{\rlap{\lower4pt\hbox{\hskip1pt$\sim$}}
     \raise1pt\hbox{$>$}}}         

\newcommand{\be}{\begin{equation}}
\newcommand{\ee}{\end{equation}}
\newcommand{\ba}{\begin{eqnarray}}
\newcommand{\ea}{\end{eqnarray}}
\newcommand{\bea}{\begin{equation}\begin{aligned}}
\newcommand{\eea}{\end{aligned}\end{equation}}
\newcommand{\baa}{\begin{array}}
\newcommand{\eaa}{\end{array}}

\begin{document}

\title{Limits on High-Frequency Gravitational Waves in Planetary Magnetospheres}

\author{Tao Liu}
\email{taoliu@ust.hk}
\affiliation{Department of Physics and Jockey Club Institute for Advanced Study, The Hong Kong University of Science and Technology, Hong Kong S.A.R., P.R.China}

\author{Jing Ren}
\email{renjing@ihep.ac.cn}
\affiliation{Institute of High Energy Physics, Chinese Academy of Sciences, Beijing 100049, P.R.China}

\author{Chen Zhang}
\email{iasczhang@ust.hk}
\affiliation{Department of Physics and Jockey Club Institute for Advanced Study, The Hong Kong University of Science and Technology, Hong Kong S.A.R., P.R.China}

\begin{abstract}

High-frequency gravitational waves (HFGWs) carry a wealth of information on the early Universe with a tiny comoving horizon and astronomical objects of small scale but with dense energy. We demonstrate that the nearby planets, such as Earth and Jupiter, can be utilized as a laboratory for detecting the HFGWs. These GWs are then expected to convert to signal photons in the planetary magnetosphere, across the frequency band of astronomical observation. As a proof of concept, we present the first limits from the existing low-Earth-orbit satellite for specific frequency bands and project the sensitivities for the future more-dedicated detections. The first limits from Juno, the latest mission orbiting Jupiter, are also presented. Attributed to the long path of effective GW-photon conversion and the wide angular distribution of signal flux, we find that these limits are highly encouraging, for a broad frequency range including a large portion unexplored before.

\end{abstract}

\maketitle

\section{Introduction}

The successful detection of gravitational waves (GWs) by the Laser Interferometer Gravitational-Wave Observatory (LIGO) opens up a window to observe the Universe otherwise inaccessible~\cite{LIGOScientific:2016aoc}. This has motivated a series of ongoing and to-be-launched projects to detect the GWs with a frequency ranging  from $\sim 10^3\,$Hz to orders of magnitude below that. Yet, the GWs with a frequency above that could have also been produced in the early cosmological events such as preheating and high-temperature phase transition and the violent astronomical activities of small-scale objects, e.g., merging of primordial black holes and intercommutation of cosmic strings. Thus, detecting high-frequency GWs (HFGWs) is of high scientific value (for a review, see, e.g.,~\cite{Aggarwal:2020olq}). 

However, the detection of HFGWs has been significantly less explored than that  of low-frequency GWs~\cite{2015RPPh...78l4901L,Tiburzi_2018,Renzini:2022alw}. Due to the shorter wavelength of HFGWs, this task is more challenging. One traditional wisdom is to employ the inverse Gertsenshtein effect~\cite{Gertsenshtein, Lupanov, Boccaletti, Zeldovich, DeLogi:1977qe, Raffelt:1987im}, where the HFGWs are expected to convert to signal photons in an astronomical~\cite{Chen:1994ch, Cillis:1996qy, Pshirkov:2009sf, Domcke:2020yzq,Ramazanov:2023nxz} or artificial~\cite{Berlin:2021txa, Aguiar:2010kn,Harry:1996gh,Herman:2020wao, Herman:2022fau, Domcke:2022rgu,Ringwald:2020ist, Ejlli:2019bqj, Li:2000du, Li:2003tv, Li:2004df, Li:2006sx, Li:2008qr, Tong:2008rz, Stephenson:2009zz, Li:2009zzy, Li:2011zzl, Li:2013fna, Li:2014bma, Li:2015nti,Li:2023tzw, Berlin:2023grv} magnetic field. To compensate for the weakness of gravitational coupling, the magnetic field needs to be either strong or distribute broadly in space. Nonetheless, the existing proposals are subject to a variety of weakness, such as relatively short path for high-efficient conversion (e.g., neutron star (NS)~\cite{Raffelt:1987im}), large uncertainty of cosmic magnetic field strength~\cite{Domcke:2020yzq}, highly specific detection frequency band (for a recent effort to address this, see~\cite{Berlin:2023grv}), and narrow angular distribution of signal flux (especially for some laboratory experiments).   

Alternatively, in this Letter, we propose to detect the HFGWs using the nearby planets such as Earth and Jupiter as a laboratory, where
the GW-photon conversion is expected to occur in their planetary magnetosphere. Due to its relatively big size, the path for the effective conversion in such a laboratory is typically long. Particularly, as to be shown, such an effective conversion can be achieved across the full electromagnetic (EM) frequency band of astronomical observation, ranging from radio waves to PeV photons. Moreover, as the detectors are positioned in the planetary magnetosphere, the stochastic signals can be detected in a wide range of directions. Combining these features creates a new operation space for detecting the HFGWs (for applying the planets to detect dark matter, see, e.g.,~\cite{Zioutas:1998ra, Davoudiasl:2005nh, Davoudiasl:2008fy, Feng:2015hja, Leane:2021tjj, Li:2022wix, French:2022ccb}). 

As a proof of concept, we consider the satellite-based detectors at low Earth orbit (LEO), with a bird view to the dark side of Earth. Both diffuse sky background and sunshine are expected to be occulted by the Earth then. We will present the first limits for some specific frequency bands and project the sensitivities for the future more-dedicated detection. It is important to note that the variety of detector designs, such as terrestrial versus satellite-based, bird view versus bottom view, etc., can have significant impacts on the sensitivities. Therefore, this study should be considered as a starting point for more systematic exploration of the opportunities presented by such a laboratory, rather than a full demonstration of the sensitivity potential of this strategy.

\section{GW-photon Conversion Probability}

With a WKB approximation,  the inverse Gertsenshtein effect is characterized by a mixing matrix~\cite{Raffelt:1987im,Ejlli:2018hke}
(see Supplementary Materials (SM) at Sec. A for details{\color{red} ~\cite{SM_link}})
\begin{eqnarray}
\left(
  \begin{array}{cc}
    \Delta_{\gamma} & \Delta_{\rm M} \\
   \Delta_{\rm M} & 0
  \end{array}
  \right) \, . 
  \label{eq_mixing}
\end{eqnarray}
Here $\Delta_{\rm M}=\frac{1}{2}\kappa B_t$ encodes the GW-photon mixing, with $\kappa=(16\pi G)^{1/2}$ and $B_t$ being the component of external magnetic field $\bf{B}$ transverse to the GW traveling direction. 
$\Delta_\gamma\approx \Delta_{\rm vac}+\Delta_{\rm pla}$ is the  effective photon mass.
$\Delta_{\rm vac}=7\alpha \omega/(90\pi) (B_t/B_c)^2$ denotes the QED vacuum effect, with $\alpha$ the fine-structure constant, $\omega$ the angular frequency and $B_{c}=m_e^2/e$. $\Delta_{\rm pla}=-m_{\rm pla}^2/(2\omega)$ represents the plasma-mass contribution with $m_{\rm pla}^2=4\pi\alpha n_c/m_c$, where $n_c$ and $m_c$ are the number density and invariant mass of charged plasma particles. 
 By diagonalizing this mixing matrix, one can obtain the GW-photon conversion probability in a homogeneous magnetic field~\cite{Irastorza:2018dyq, Domcke:2020yzq}
\ba
P
=\sin^2(2\Theta) \sin^2\left ( \frac{L}{l_{\rm osc}} \right)
=(\Delta_{\rm M} L)^2  \textrm{sinc}^2\left ( \frac{L}{l_{\rm osc}} \right)
\,.   \label{PhomoB}
\ea
Here $\Theta=\frac{1}{2} \arcsin(\Delta_{\rm M} l_{\rm osc})$ and $l_{\rm osc}=2/(4\Delta_{\rm M}^2+\Delta_{\gamma}^2)^{1/2}$ are the GW-photon mixing angle and oscillation length, respectively. $L$ is the travel distance of GWs in the magnetic field. 
For a general path from $\ell_0$ to $\ell_1$, the conversion probability can be evaluated as~\cite{Raffelt:1987im, footnoteEq3}:
\be
P={\bigg| \int_{\ell_0}^{\ell_1} d\ell~\Delta_{\rm M}(\ell) 
\exp \left( -i \int_{\ell_0}^\ell d\ell'~\Delta_{\gamma} (\ell') \right) \bigg|}^2\,.
\label{pint}
\end{equation}

\begin{figure}[h]
\centering
\includegraphics[width=8cm]{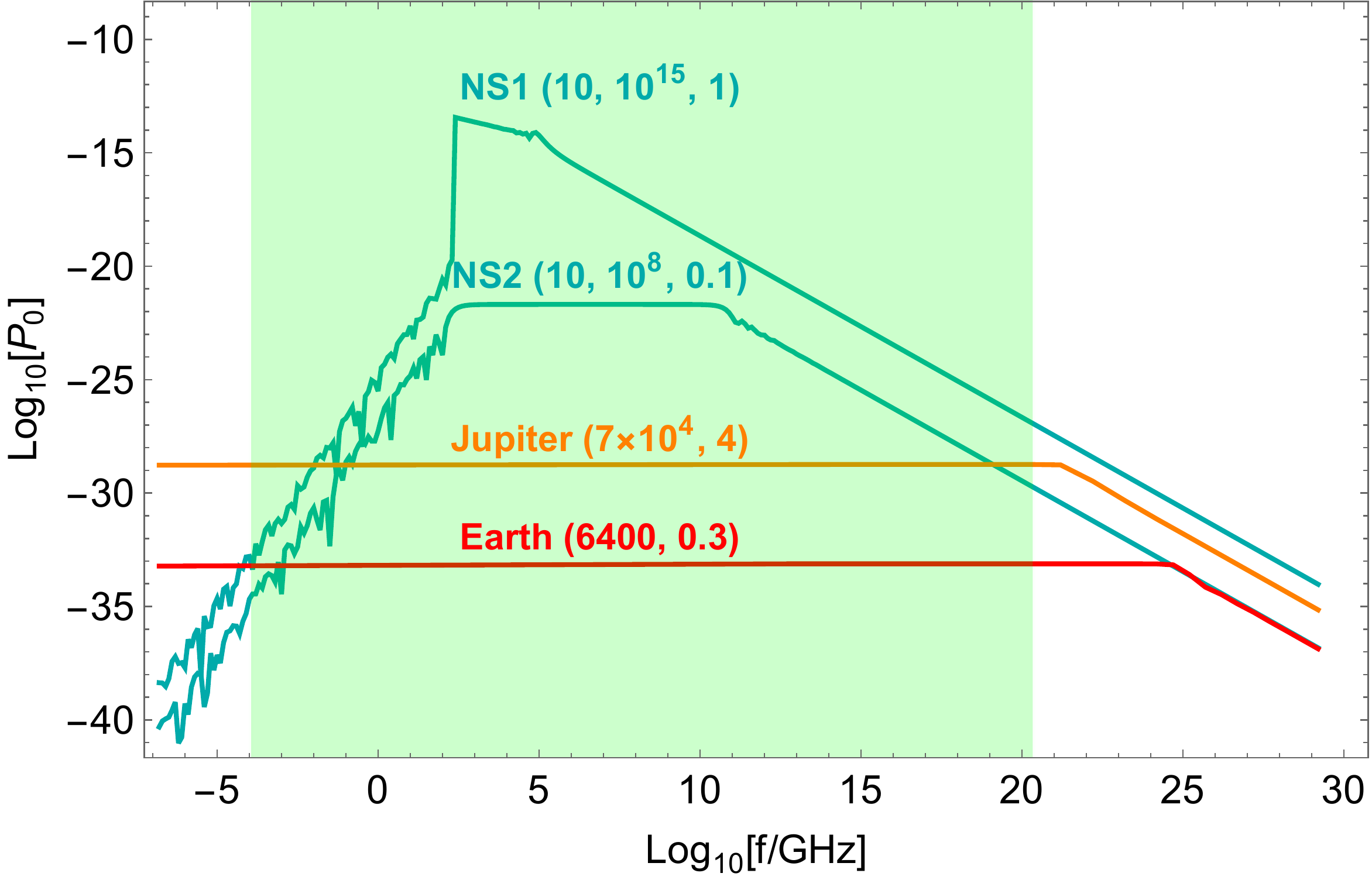}
\caption{$P_0$ as a function of $f=\omega/2\pi$. 
We show $(r_0,B_0)$ for the planets and $(r_0,B_0,2\pi/\Omega_{\rm NS})$ for the NSs, where $r_0$, $B_0$ and $\Omega_{\rm NS}$ (spinning velocity) are in unit of km, Gauss and second$^{-1}$, respectively.  
The green region denotes the EM observation frequency band in astronomy, ranging from radio waves to PeV photons.}
\label{fig:p_all}
\end{figure}

To have a taste, we first evaluate $P_0$, the conversion probability for a radial path from zero altitude to infinity over  the planet equator. We consider the NSs also for reference. The magnetic fields of both can be modeled as a magnetic dipole, with the magnetic axis aligned with the rotation axis. Then they are transverse to the radial path, with $B_t=-B_0(r_0/r)^3$. Here $r_0$ is radius and $B_0$ is surface magnetic field strength. We show $P_0$ as a function of frequency for the Earth, Jupiter and two benchmark NSs in Fig.~\ref{fig:p_all}. Because of their difference in plasma density profile and external magnetic field strength, the curves demonstrate quite different features.  

For the planets, the plasma density is described by a barometric formula, which yields an exponentially suppresses plasma mass as $r$ increases. If $\omega$ is not too small, we can set $\Delta_{\gamma} \approx  \Delta_{\rm vac}$ and then obtain
\ba
P_0= \left\{
\begin{array}{ll}
\frac{1}{4}\Delta^2_{\rm M}(r_0)\, r^2_0\propto B_0^2 r_0^2 \, ,  &  \omega\lesssim \omega_{\rm tra} \\
\frac{1}{4}\Delta^2_{\rm M}(r_0)\, r^2_0 \left[\frac{\omega}{\omega_{\rm tra}}\right]^{-\frac{4}{5}}\propto B_0^{\frac{2}{5}} r_0^{\frac{6}{5}}\omega^{-\frac{4}{5}}, &  \omega\gtrsim \omega_{\rm tra} \\
\end{array}\right.
\label{eq:Papprox}
\ea
where $\omega_{\rm tra}$ is 
determined by $r_0\approx l_{\rm osc}(r_0)$,  corresponding to the transition point of planet curves in Fig.~\ref{fig:p_all}.  Consisting with Fig.~1 which is full-calculation-based, $P_0$ is approximately a constant for $\omega\lesssim \omega_{\rm tra}$ and drops as $\propto \omega^{-4/5}$ for $\omega\gtrsim \omega_{\rm tra}$.
These relations can be qualitatively explained with Eq.~(\ref{PhomoB}). In the homogeneous magnetic field, an optimal probability $P \approx \Delta_{\rm M}^2 L^2$ can be achieved for $ \textrm{sinc} (L/l_{\rm osc}) \to 1$ or $l_{\rm osc} \gtrsim L$, a case dubbed as ``coherent conversion''~\cite{Mirizzi:2009aj,Kartavtsev:2016doq}, for given $L$. When $l_{\rm osc}$ becomes smaller than $L$, $P$ is suppressed by $l_{\rm osc}^2/L^2$. Extending this criterion to the planets, we have coherent conversion near their surface for $L \sim r_0\lesssim l_{\rm osc}(r_0)$ or $\omega \lesssim \omega_{\rm tra}$, where $P_0 \propto \Delta^2_{\rm M}(r_0) r_0^2 \propto B_0^2 r_0^2$. $P_0$ for the Jupiter in this region is then $\sim 10^4$ times larger than that of the Earth, as its $B_0$ and $r_0$ are both ten times larger.  
For $r_0 > l_{\rm osc}(r_0)$, the coherent conversion is suppressed near the planet surface.  
It can only occur for $r_* \approx l_{\rm osc}(r_*)> r_0$, with a reduced rate $P_0 \sim \Delta^2_{\rm M}(r_*) r_*^{2} \propto B_0^{2/5} r_0^{6/5}\omega^{-4/5}$. As shown in Fig.~\ref{fig:p_all}, the frequency band for EM astronomical observations falls entirely into the range of near-surface coherent conversion for both Earth and Jupiter. 

In comparison, due to the large strength of their external magnetic field, the NSs tend to have a suppressed $l_{\rm osc}$. The near-surface coherent conversion thus becomes difficult to achieve. For the two benchmark NSs in Fig.~\ref{fig:p_all}, it takes place between $10^2 - 10^{11}\,$GHz for the NS2, and hardly occurs for the NS1~\cite{NS1}. 
In the high-frequency limit, the vacuum effect dominates. $P_0$ is given by the 2nd formula in Eq.~(\ref{eq:Papprox}). At the low-frequency end, the plasma effect becomes dominant, yielding a sine-wiggling $P_0$. With the Goldreich-Julian model for the NS plasma density~\cite{Goldreich:1969sb}, where $|\Delta_{\rm pla}| \propto n_c = \frac{2}{e}\mathbf{\Omega}_{\rm NS} \cdot \mathbf{B} [1 - (\Omega_{\rm NS}\, r/c)^2 \sin^2 \theta]^{-1} \propto {|\bf B|}$, $|\Delta_{\rm pla}|$ decays more slowly than $\Delta_{\rm vac}$ does as $r$ increases. The coherent conversion then takes place at a larger $r_*$, compared to the high-frequency case, yielding a more suppressed $P_0$ as $\omega$ decreases~\cite{LRZ}.

\begin{figure}[h]
\centering
\includegraphics[width=8cm]{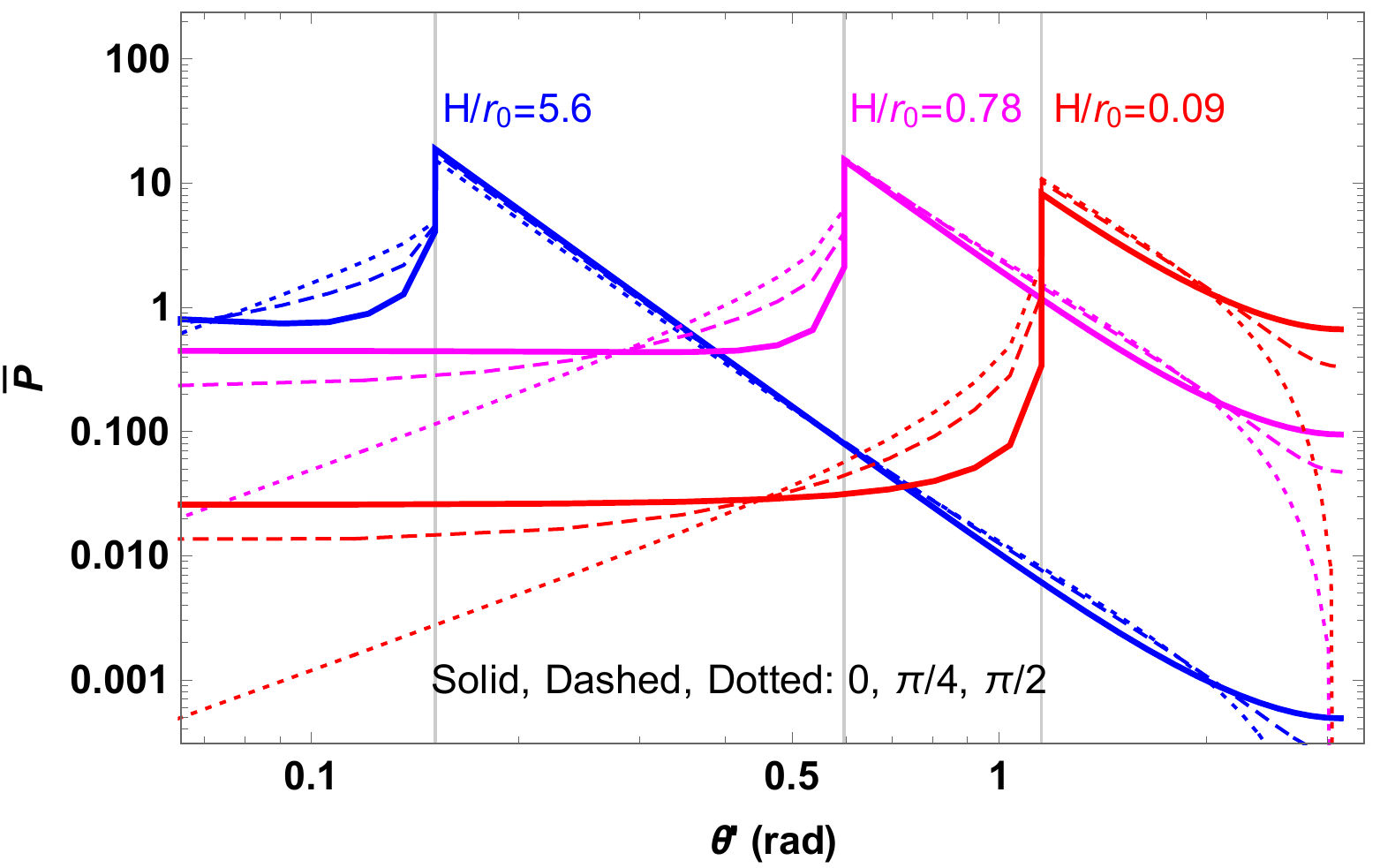}
\caption{$\bar{P}(\theta')$ as a function of $\theta'$ for a satellite-based detector, with $f=10^8\,$GHz. The color and line style denote the normalized altitude $H/r_0$ and the latitudes $\lambda_{O'}$ respectively for the satellite. 
The vertical lines denote $\theta_c$ for the PC with different $H/r_0$ values. } 
\label{fig:p_thetaL}
\end{figure}

Now let us consider a satellite-based detector positioned at altitude $H$ and latitude $\lambda_{O'}$, and define its instant spherical coordinate system with ${\bf \hat z}'$ pointing to the planet's center (see SM at Sec.~A for details). 
Then we are able to evaluate $P(\Omega')$ of HFGWs traveling to this detector in all directions, where $\Omega'=\{\theta',\phi'\}$.  
Fig.~\ref{fig:p_thetaL} displays the $\theta'$ dependence of $P(\Omega')$, with $\bar{P}(\theta') =(2\pi P_0)^{-1}\int_0^{2\pi} P(\Omega')d\phi'$. While this figure is drawn with $f=10^8\,$GHz, the features that it demonstrates are almost unchanged for the GW frequency range of interest.
Here we neglect the plasma effect which is expected to be small. We split the photon incoming directions into planet-cone (PC) ($\theta'<\theta_c= \arcsin [r_0/(r_0+H)]$) and outer-space (OS) ($\theta'>\theta_c$) regions, based on whether the line of sight intersects with planet surface. For all sets of $(H/r_0, \lambda_{O'})$, $\bar{P}(\theta')$ peaks sharply at the PC edge and drops quickly away from it. 
For $\lambda_{O'}=\pi/2$, it drops all the way to zero at $\theta'=0$, as $B_t$ vanishes in this direction for a detector right above the planet poles. 
As $H/r_0$ increases, with a cost of smaller PC, $\bar{P} (\theta')$ tends to have a bigger value inside the PC due to a longer GW-photon conversion path.

\section{Sensitivity Analysis}

The stochastic GWs are typically isotropic and stationary~\cite{Romano:2016dpx}. For a detector in the planetary magnetosphere, we have the GW-converted photon flux (see, e.g.,~\cite{Leroy:2019ghm})    

\begin{eqnarray}\label{eq:fluxg}
\Phi_\gamma=\int_{\Delta \Omega} d\Omega' \int d \omega  \frac{1}{\omega}\frac{d}{d\omega}\frac{d \rho_{\rm GW}}{d\Omega} P(\Omega') \,,
\end{eqnarray}
where $\Delta\Omega$ denotes the detector field of view (FOV), $d^2 \rho_{\rm GW}/d\Omega d\ln\omega=\omega^2h_c^2/(4\pi\kappa^2)$~\cite{Romano:2016dpx} and $h_c$ is the GW characteristic strain. 
Note, the angular distribution of GW-converted photons is determined by $P(\Omega')$ and only those falling into the detector FOV can contribute to $\Phi_\gamma$.
The signal and background counts for a narrow frequency band $\Delta \omega$, a short observation time $\Delta t$ and an effective detector area $A$ are then given by
\begin{eqnarray}\label{eq:DeltaSB}
 s&\approx&\Phi_\gamma A\, \Delta t 
\approx  \frac{h_c^2}{4\pi \kappa^2}  \langle P \rangle_{\rm det}   A\,  \Delta t \Delta \omega 
\Delta\Omega   \, , 
\nonumber\\
 b&\approx& \Phi_b A\, \Delta t   \approx  \phi_b  A\,\Delta t \Delta\omega \Delta \Omega\,.
\end{eqnarray} 
Here $\langle P \rangle_{\rm det} = \int_{\Delta\Omega} P (\Omega') d\Omega' /\Delta \Omega$, essentially determined by the detector position and pointing direction \{$\theta’_{\rm det}, \phi’_{\rm det}$\}, denotes the average GW-photon conversion probability over its FOV. If the detector points to the planet center ($\theta’_{\rm det}=0$), $\langle P \rangle_{\rm det}$ is $\propto \int_{\Delta\Omega} \bar P (\theta') \sin \theta' d\theta' / \int_{\Delta\Omega} \sin \theta' d\theta'$ and increases with $\Delta\Omega$ inside the PC. The 95\% upper limit on $h_c$ can be derived from $s/\sqrt{b}\approx 1.64$ in the large-background limit: 
\begin{eqnarray}
\resizebox{0.43\textwidth}{!}{$
\begin{aligned}
h_{c, 95\%}&\approx 4.5 \kappa\left(\frac{\phi_b}{A\, \Delta t \Delta \omega \Delta \Omega}\right)^{1/4}\left(\frac{1}{\langle P \rangle_{\rm det}}\right)^{1/2}.
\end{aligned}
$}
\label{eq:hcUL1}
\end{eqnarray}

In this Letter, we mainly consider the LEO satellite which has a bird view to the dark side of Earth.  With the detector FOV restricted to inside the PC, the dominant backgrounds vary from atmospheric thermal emissions in the infrared (IR) band~\cite{B294K} to cosmic photon albedo (see, e.g.,~\cite{Turler:2010pm}) throughout the optical - $\gamma$-ray region. We simplify the analysis by assuming a uniform background flux $\phi_b$. 
We will briefly discuss the Jupiter case also.  
Given the strong angular dependence of $P(\Omega')$ (see Fig.~\ref{fig:p_thetaL}), an efficient detection requires proper design for satellite orbit and full optimization of detector performance. Below let us consider some specific examples.

\begin{table*}
\begin{center}
\scalebox{1}{
\begin{tabular}{c||ccc||c|ccccc}
\hline\hline
&  \multicolumn{3}{c||}{Satellite orbit}
     &  
     \multicolumn{6}{c}{Detector properties} \\
\cline{2-10}
     &  $H$\,(km)
     &  $\theta_{\rm inc}$ 
     & $T_{\rm dark}$\,(s)
     &    
     &  IR
     &  UV-Optical
     &   EUV
     &  X-ray
     & $\gamma$-ray  \\
\hline
\multirow{2}{*}{Conservative }
& \multirow{2}{*}{600} 
& \multirow{2}{*}{$31.4^{\circ}$}
& \multirow{2}{*}{$10^7$}
& $\Delta \Omega$ (sr)\,\, & \,\,$1.6\times 10^{-2}$\,\, & \,\, $10^{-6}$\,\, & \,\,   $10^{-5}$  \,\,  & \,\,$3\times10^{-5}$\,\, &  \,\,$2.4$\,\,
  \\
 &&&& $A$ ($\rm cm^2$)\,\, & \,\,0.1225\,\, & \,\, $4.5\times10^4$\,\, & \,\,  1  \,\, & \,\,250\,\, &  \,\,8000\,\,\\
\hline
 \multirow{2}{*}{Optimistic } 
& \multirow{2}{*}{800} 
& \multirow{2}{*}{$98^{\circ}$} 
& \multirow{2}{*}{$10^8$} 
& $\Delta \Omega$ $(\rm sr)$\,\, & \,\,3.4\,\, & \,\, 3.4\,\,  & \,\,  3.4  \,\, & \,\,  3.4  \,\, &  \,\,3.4\,\,  \\
&&&& $A$ ($\rm cm^2$)\,\, & \,\,0.1\,\, & \,\, $10^2$\,\,  & \,\,  $10^2$  \,\, & \,\,$10^2$\,\, &  \,\,$10^4$\,\,   \\
\hline\hline
\end{tabular}}
\caption{ 
Benchmark scenarios for sensitivity study. For the conservative case, we take the Suzaku-like low inclination orbit~\cite{Suzaku} with one year operation. The  detector FOV and effective area are assumed to be the same as those of the existing missions including Nimbus~\cite{hanel1965infrared} (IR), Hubble~\cite{Hubble1, Hubble2} (ultraviolet (UV)-Optical), Voyager~\cite{broadfoot1977ultraviolet} (extreme UV (EUV)), Suzaku~\cite{koyama2007x} (X-ray) and Fermi-LAT~\cite{Fermi} ($\gamma-$ray). For the optimistic case, we take the Safir-2-like high inclination orbit~\cite{Safir2} with ten year observation. A FOV covering the whole PC is considered. The effective area is set in a way such that the corresponding etendue for the IR, UV-optical, EUV and X-ray bands~\cite{forster2021automatic} is consistent with that of future missions~\cite{TESS,Law_2015,EP,VLAST}. 
}
\label{tab:tele}
\end{center}  \
\end{table*}

We first consider the detection of HFGWs in the Earth's magnetosphere. Take the Suzaku mission~\cite{koyama2007x} as an example. The Suzaku has a low inclination orbit (see Tab.~\ref{tab:tele}). It revolves at low latitude and has a performance relatively insensitive to season alternation. Eq.~(\ref{eq:hcUL1}) thus applies approximately for the entire observation period $T_{\rm dark}$. 
Also, as the Suzaku FOV is small, $\langle P \rangle_{\rm det}$ does not vary much for $\theta’_{\rm det} \ll \theta_c$.  For $\theta’_{\rm det}=0$ we have $\langle P \rangle_{\rm det} \approx  5 \times 10^{-35}$. 
As for the background flux in the dark side, it has been measured by the Suzaku to be  $\phi_b \approx 6.3\,\times 10^{-8}\textrm{cm}^{-2}\textrm{s}^{-1}\textrm{keV}^{-1}\textrm{arcmin}^{-2}$ for $\omega \sim 0.5-10\,$keV~\cite{koyama2007x}. 
For one-year operation, we have 
\begin{eqnarray}\label{eq:hcSuzaku}
h_{c,95\%}&\sim & 2 \times 10^{-25} \, .
\label{sim_hc}
\end{eqnarray}
This limit can be scaled to other missions with similar properties, following  Eq.~(\ref{eq:hcUL1}).

Notably, the satellite at a high inclination orbit has quite different properties. It scans over the high latitude region also, where the conversion probability varies more inside the PC and becomes bigger for a region extending from the PC edge to its inside, compared to the low-latitude case (see Fig.~\ref{fig:p_thetaL}). The sensitivity thus could be optimized by taking a detector with either a large FOV covering the whole PC or a small FOV but pointing to the region near the PC edge. Moreover, the observation of such a satellite in the dark side of Earth is sensitive to the season alternation. So we need to generalize Eq.~(\ref{eq:hcUL1}) by binning the detector FOV and observation time in this case. The limit shall be obtained from the combined statistics (see SM at Sec.~B for details).

Next, we consider the HFGWs conversion in the Jovian magnetosphere. Take Juno, the latest mission orbiting the giant, as an example. The Juno is settled at a highly elliptical polar orbit with a Perijove $\sim 5000\,$km~\cite{2013AGUFMGP54A..03B} and a high inclination angle ($\sim 90^{\circ}$). 
For each orbit, the Juno spends a few hours around the Perijove in observing the  Jupiter's cloud. So $\Delta t$ is $\sim 10^5\,$s for the $\sim 35$ rounds of its prime mission. As a conservative estimate, we consider the Juno observation of aurora emissions by the Jovian Infrared Auroral Mapper (JIRAM)~\cite{Adriani2017} and  UV emissions by the Ultraviolet Spectrograph (UVS)~\cite{2017SSRv..213..447G}, which set respectively $\phi_b \sim 4.2\times 10^{4}\, \textrm{cm}^{-2}\textrm{s}^{-1}\textrm{eV}^{-1}\textrm{arcmin}^{-2}$ for $\omega\sim 0.35 \,$eV~\cite{2017GeoRL..44.4633A} and $ \sim 34\, \textrm{cm}^{-2}\textrm{s}^{-1}\textrm{eV}^{-1}\textrm{arcmin}^{-2}$ for $\omega\sim 8 \,$eV~\cite{Giles_2023}. We take $\theta'_{\rm det}\sim 0.8\theta_c$ for $\langle P\rangle_{\rm det}$ and assume $\langle P\rangle_{\rm det}$ to vary little with time. 
Then, with $\Delta \Omega \sim 10^{-3}\,$sr and $3\times10^{-4}$ sr, $A\sim 2\,\text{cm}^2$ and $6\,\text{cm}^2$ and $\Delta \omega \sim 1.4\times 10^{13}\,$Hz and $0.9\times 10^{15}\,$Hz, we have
\begin{eqnarray}
h_{c,95\%}\sim 4 \times 10^{-22} \quad {\rm  and} \quad  2 \times 10^{-23} \, ,
\end{eqnarray}
for the JIRAM and the UVS, respectively.

As a reference, let us consider the conversion of HFGWs in the NS magnetosphere. The NSs are remote. So we have $\langle P\rangle_{\rm det}\sim P_0$ 
and $ \Delta\Omega\sim \pi r_0^2/d^2$, where $d$ is their distance to the Earth.
Take the Magnificent Seven (M7) X-ray dim isolated NSs as an example, which have $d\sim \mathcal{O}(100)\,$pc and $B_0\sim 10^{13}\,$Gauss, and hence $P_0\sim 10^{-18}$ and $\Delta\Omega\sim 10^{-30}\,\textrm{sr}$.
The PN and MOS of the XMM-Newton telescope have measured the flux of, e.g., J0420 ($d\approx 345\,$pc), to be $\phi_b \sim 10^{17}-10^{19}\, \textrm{cm}^{-2}\textrm{s}^{-1}\textrm{keV}^{-1}\textrm{arcmin}^{-2}$ for  $\omega \sim 0.5-1\,$keV, where $A \sim 1500\rm \, cm^2$ and $\Delta t \sim 10^5\,$s. Such an intensity consists well with the background model of thermal surface emissions~\cite{Dessert_2020}.
By applying Eq.~(\ref{eq:hcUL1}), we find  
\begin{eqnarray}
h_{c,95\%}&\sim& \mathcal O(1) \times 10^{-20} \, .
\label{NSs}
\end{eqnarray}
Due to the extreme smallness of $\Delta\Omega$, this limit is about five orders of magnitude worse than that in Eq.~(\ref{eq:hcSuzaku}). This highlights in part the merit of nearby planets in performing the task of detecting HFGWs.

It is also informative to compare these limits with the ones recast from the existing laboratory experiments for axion detection~\cite{Ejlli:2019bqj}. In these experiments, the magnetic field is typically confined inside a long straight pipe, with a small cross-sectional area. The GW-photon conversion is suppressed outside the small opening angle of long pipe. The signal flux in Eq.~(\ref{eq:fluxg}) is thus strongly limited by the detector geometry  (see SM at Sec.~C for details). Taking the CERN Axion Solar Telescope (CAST) experiment~\cite{CAST:2017uph} as an example, we have the recast limit  
\begin{eqnarray}\label{eq:hcCAST}
h_{c,95\%}\sim 8 \times10^{-26}\,.
\end{eqnarray}

\section{Projected sensitivities}

\begin{figure}[h]
\centering
\includegraphics[width=8cm]{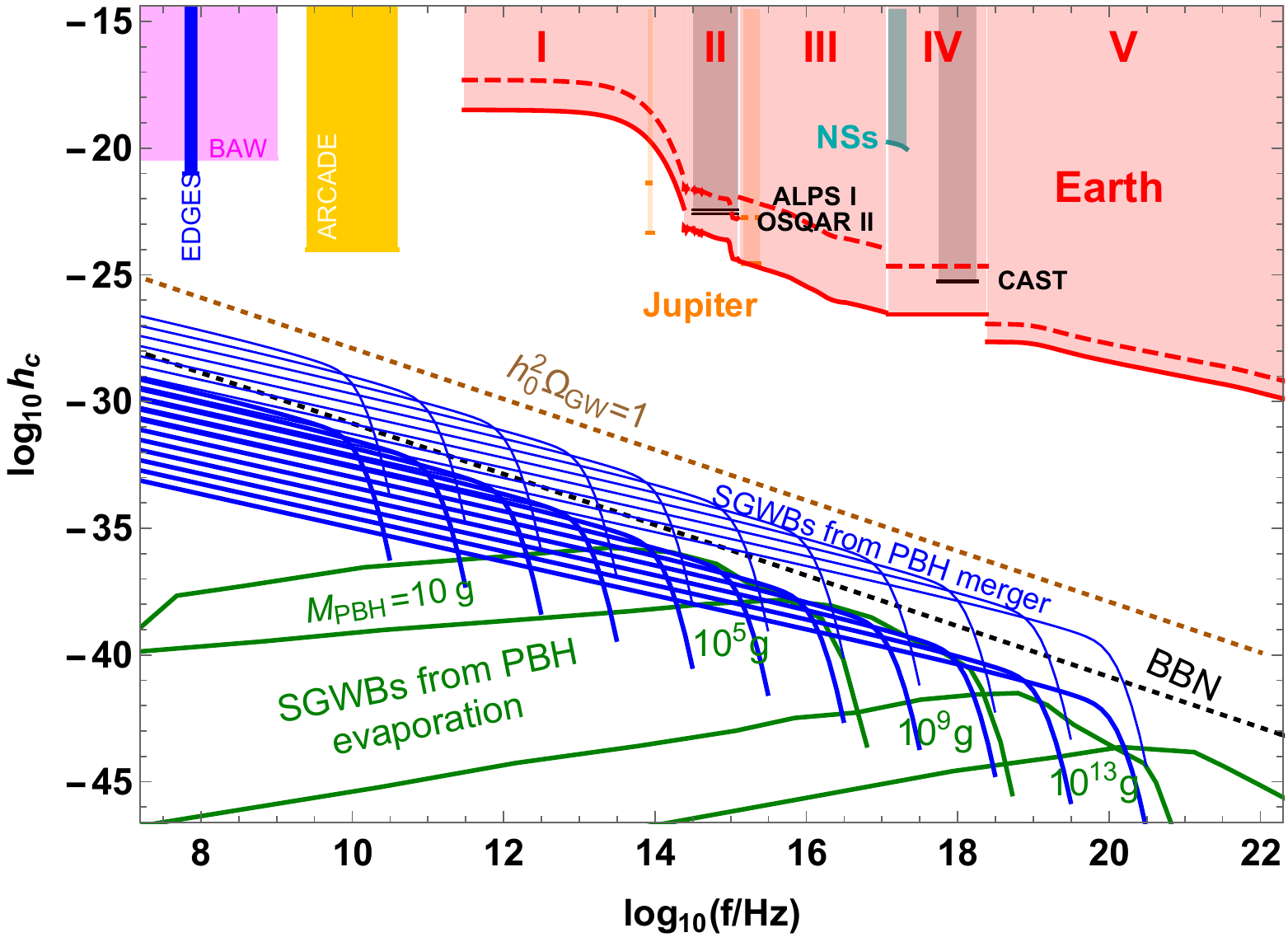}
\caption{Projected $95\%$ C.L. upper limits on the characteristic strain $h_c$ of stochastic HFGWs, set by the LEO satellites (red). We estimate these limits with the Suzaku X-ray measurements~\cite{koyama2007x} for band IV, and the  IR~\cite{B294K}, UV-optical~\cite{Hauser:2001xs,Cooray:2016jrk,7374696,tilstra2004reflectance}, EUV~\cite{Hill:2018trh,cravens2006x} and $\gamma$-ray~\cite{Cumani:2019ryv} backgrounds for bands I, II, III and V, respectively. The dashed and solid bottom lines for these bands correspond to the conservative and optimistic estimates defined in Tab.~\ref{tab:tele}.  Moreover, we present the conservative and optimistic limits as orange bands, based on the Juno observation of aurora emissions~\cite{2017GeoRL..44.4633A} and UV spectra at Jupiter~\cite{Giles_2023}. We also present the limits set by the X-ray observations of M7 NSs~\cite{Dessert_2020} as cyan band.  
For completeness, we recast the limits of ALPS, OSQAR and CAST, {\it i.e.}, three existing laboratory experiments, as gray bands. We denote the highly-uncertain radio limits of EDGES and ARCADE~\cite{Aggarwal:2020olq} as blue and yellow bands~\cite{footnote4}. The black dotted line reflects the Big-Bang-Nucleosynthesis (BBN) constraints on the radiation energy density in the early Universe~\cite{Cyburt:2015mya}, while the khaki dotted line represents the maximal strain values set by the critical energy density of the Universe today. 
As a reference, we present the HFGWs from the Hawking evaporation of highly spinning primordial black holes (PBHs) in the early Universe~\cite{Anantua:2008am,Dolgov:2011cq,Dong:2015yjs} as green solid lines, and from the binary mergers of PBHs after the BBN~\cite{Franciolini:2022htd} as thin-solid and solid blue lines for two benchmark scenarios. While the former is subject to the BBN constraints, the latter is less restricted.} 
\label{fig:limit}
\end{figure}

To demonstrate the potential of detecting stochastic HFGWs with the LEO satellites, we define two benchmark scenarios, dubbed ``conservative'' and ``optimistic'', in Tab.~\ref{tab:tele}. The choice of orbit in these two scenarios echoes the discussions above on its impacts on the detection efficiency. Fig.~\ref{fig:limit} shows the projected $95\%$ C.L. limits on the characteristic strain of HFGWs for a wide range of frequencies, including a large portion unexplored before.

In the $\gamma$-ray band, the limits are strengthened with a rate faster than $\sim f^{-1/2}$, due to the reduction of albedo backgrounds~\cite{Cumani:2019ryv}.
For the $X$-ray region, the background flux is taken from the Suzaku measurement. Its conservative limits thus represent the first limits from this mission. 
For the UV-optical band, we consider the cosmic background~\cite{Hauser:2001xs,Cooray:2016jrk} with a frequency-dependent reflectance of atmosphere~\cite{7374696,tilstra2004reflectance}.
The limits get nearly one order of magnitude stronger at the right edge of the UV-optical band, due to the ozone absorption of cosmic photons. Such a trend extends to the EUV band, where photodissociation and photoionization become important~\cite{FU2006113} and a reflectance of $\sim 10^{-3}$~\cite{cravens2006x} is thus approximately taken for cosmic photons. As for the IR band, the backgrounds are modelled with a blackbody spectrum at 294K~\cite{B294K}. 
These thermal radiations peak at $f\sim 10^{13}\,$Hz,
and are quickly suppressed for $f > 10^{14}\,$Hz. We also present the limits from the observations of Jupiter and NSs in some specific frequency bands. For the optimistic estimate of Jupiter limits, we take $\Delta \Omega = 4\,$sr, $A=100\, \text{cm}^2$ and $\Delta t =10^7\,$s.
 Notably, the energy density contribution of GWs with such a large stochastic magnitude at these frequencies is too significant, compared to the BBN constraints and the cosmic critical value today. Nevertheless, the presented limits lay a foundation for further searching for astrophysical signals of, e.g., PBH mergers and advancing detector technologies capable of detecting the signals from the early Universe.

\section{Summary and Outlook}

In this Letter, we have proposed to detect the stochastic HFGWs in the planetary magnetosphere. Due to the relatively long path for effective GW-photon conversion, the wide angular distribution of signal flux and a full coverage of the EM observation frequency band in astronomy, this strategy creates a new operation space.  With the proof of concept presented, we can immediately see several important directions for next-step explorations.

Firstly, extend the detection of HFGWs from the PC to the OS. As indicated by Fig.~\ref{fig:p_thetaL}, the GW-photon conversion probability right outside the PC can be much higher than that inside. Moreover, a detector oriented toward the OS may receive considerably less atmospheric thermal radiation. 
Secondly, extend the satellite-based detection to terrestrial observations, which may allow the sensitivities to cover the radio bands.
Thirdly, extend the dedicated study to the Jupiter and even the Sun, given their stronger magnetic field, larger space for an effective GW-photon conversion, and the active and upcoming missions. We leave these explorations with refined analysis to a paper in preparation~\cite{LRZ}.

\begin{acknowledgments}
\subsection*{Acknowledgments}
We would greatly thank Lian Tao for discussion on the properties of satellites and X-ray telescopes and A. Ejlli for communications on the recast limits of laboratory experiments. 
T.~Liu and C. Zhang are supported by the Collaborative Research Fund under Grant No. C6017-20G which is issued by Research Grants Council of Hong Kong S. A. R.
J. Ren is supported by the Institute of High Energy Physics, Chinese Academy of Sciences, under Contract No. Y9291220K2. 

All authors contributed equally to this work.
\end{acknowledgments}

\bibliography{References}

\begin{thebibliography}{100}
\bibitem[{NS1()}]{NS1}
\bibinfo{note}{For NS1 with $f\sim 10^2-10^5\,$GHz, the dominant contribution
  to $P_0$ arises from the integration over a specific tiny range of $r$, where
  a maximal mixing is achieved with a strong cancellation between $\Delta_{\rm
  vac}$ and $\Delta_{\rm pla}$ in $\Delta_\gamma$.}
\bibitem[{foo({\natexlab{b}})}]{footnote4}
\bibinfo{note}{As the strength of cosmic magnetic field is highly
  unconstrained, these radio limits are subject to a variation of orders of
  magnitude. While the strongest ones are presented in this figure, it is
  informative to know that the weakest ones are even weaker than $10^{-14}$.}
\end{thebibliography}

\clearpage
\newpage
\maketitle
\begin{center}
\textbf{\large Limits on High-Frequency Gravitational Waves in Planetary Magnetospheres} \\ 
\vspace{0.05in}
{ \it \large Supplementary Material}\\ 
\vspace{0.05in}
{}
\end{center}
 \maketitle
The Supplementary Materials contain additional calculations and explanations in support of the
results presented in this Letter. Concretely, we discuss the GW-photon conversion in planetary magnetosphere in Sec.~A, analyze the impacts of the orbit design for the LEO satellites on the HFGWs detection in Sec.~B, and calculate the recast limits from the laboratory experiments in Sec.~C. 

\subsection{A. GW-Photon Conversion in Planetary Magnetosphere}
\label{app:conversion}

As gravity couples with EM dynamics via $\sim g^{\mu\nu} T_{\mu\nu}^{\rm EM}$, the GWs can be converted to photons while propagating in an external magnetic field $\mathbf{B}$. 
With a WKB approximation, which is valid here since the wavelength for the GW-converted photons expected to observe through astronomical telescopes is much shorter than the characteristic spatial scale (comparable to the planet size) at which the external magnetic field varies,  
the leading-order equations of motion are given by~\cite{Raffelt:1987im,  Ejlli:2018hke}  
\begin{eqnarray}
 \left( \omega-i \partial_n+
 \left( \begin{array}{cccc} 
 \Delta_\perp & \Delta_M & \Delta_R & 0 \\ 
 \Delta_M & 0 & 0 & 0 \\ 
 \Delta_R & 0 & \Delta_{||} & \Delta_M\\ 
 0 & 0 & \Delta_M & 0 \end{array} \right)
 \right)
 \left( \begin{array}{c} A_{||}\\ G_{\times}\\ A_{\perp}\\ G_{+} \end{array} \right)=0 \,.\nonumber\\
  \label{eq_mixing0}
\end{eqnarray}
Here $G_{\times}$ and $G_{+}$ denote the cross and plus polarization modes of GWs; $A_{||}$ and $A_{\perp}$ denote the polarization modes of photons parallel with and perpendicular to $\mathbf{B_T}$, the $\mathbf{B}$ component transverse to the GW propagation direction. In the $4\times 4$ propagation matrix, 
\begin{eqnarray}
\Delta_{||}&=&\Delta_\textrm{pla}+\Delta_{\textrm{vac},||}+\Delta_\textrm{CM} \,  \nonumber \\
\Delta_{\perp}&=&\Delta_\textrm{pla}+\Delta_{\textrm{vac},\perp} \, 
\end{eqnarray}
denote an effective mass of $A_{||}$ and $A_{\perp}$, respectively. Here 
\begin{eqnarray}
\Delta_{\rm pla}&=&-\frac{m_{\rm pla}^2}{2\omega}   \, , \nonumber \\
\Delta_{\textrm{vac},|| (\perp)}&=&7(4)\frac{\alpha \omega}{90\pi} \left (\frac{B_t}{B_c}\right)^2   
\end{eqnarray}
encode the plasma and QED vacuum effects, and $\Delta_{\rm CM} \propto B_t^2$~\cite{Ejlli:2018hke} can induce Cotton-Mouton birefringence. $m_{\rm pla}^2=4\pi\alpha n_c/m_c$ is plasma mass square, with $n_c$ and $m_c$ being the number density and invariant mass of charged plasma particles.
Moreover, the $\Delta_R$ term, which is proportional to $\mathbf{B} - \mathbf{B_T}$, couples $A_{||}$ and $A_{\perp}$ and can yield Faraday rotation. 
The GW-photon mixing, which is a core of this study, is governed by $\Delta_{\rm M}=\frac{1}{2}\kappa B_t$, where $B_t = |\mathbf{B_T}|$.

For the GW-photon conversion in planetary magnetosphere where $\omega$ of interest (see Fig.~\ref{fig:limit}) is mostly larger than the cyclotron angular frequency $\omega_c\approx 10^7(|\mathbf{B}|/\rm Gauss)$\,Hz, $\Delta_{\rm CM}$ and $\Delta_R$ are negligibly small compared to $\Delta_{\rm pla}$~\cite{Ejlli:2018hke}. 
Thus, the equations in Eq.~(\ref{eq_mixing0}) can be separated to two decoupled ones, for the polarization modes $\{A_{||}, G_{\times}\}$ and $\{A_{\perp}, G_{+}\}$, respectively.  
For simplicity, one can further neglect the small difference between $\Delta_{\textrm{vac},||}$ and $\Delta_{\textrm{vac},\perp}$, and assume $\Delta_{\textrm{vac},||}= \Delta_{\textrm{vac},\perp} = \Delta_{\textrm{vac}}$. The two propagating equations then share the same form and become    
\begin{eqnarray}
 \left( \omega-i \partial_n+\left(
  \begin{array}{cc}
    \Delta_{\gamma} & \Delta_{\rm M} \\
   \Delta_{\rm M} & 0
  \end{array}
  \right)\right)
  \left( \begin{array}{cc} A_{||(\perp)}\\ G_{\times(+)}\\ \end{array} \right)=0 \, ,
  \label{eq_mixing1}
\end{eqnarray}
where $\Delta_\gamma\approx \Delta_{\rm vac}+\Delta_{\rm pla}$. This gives exactly the mixing matrix in Eq.~(\ref{eq_mixing}).

\begin{figure}[h]
 \centering
\includegraphics[width=8cm]{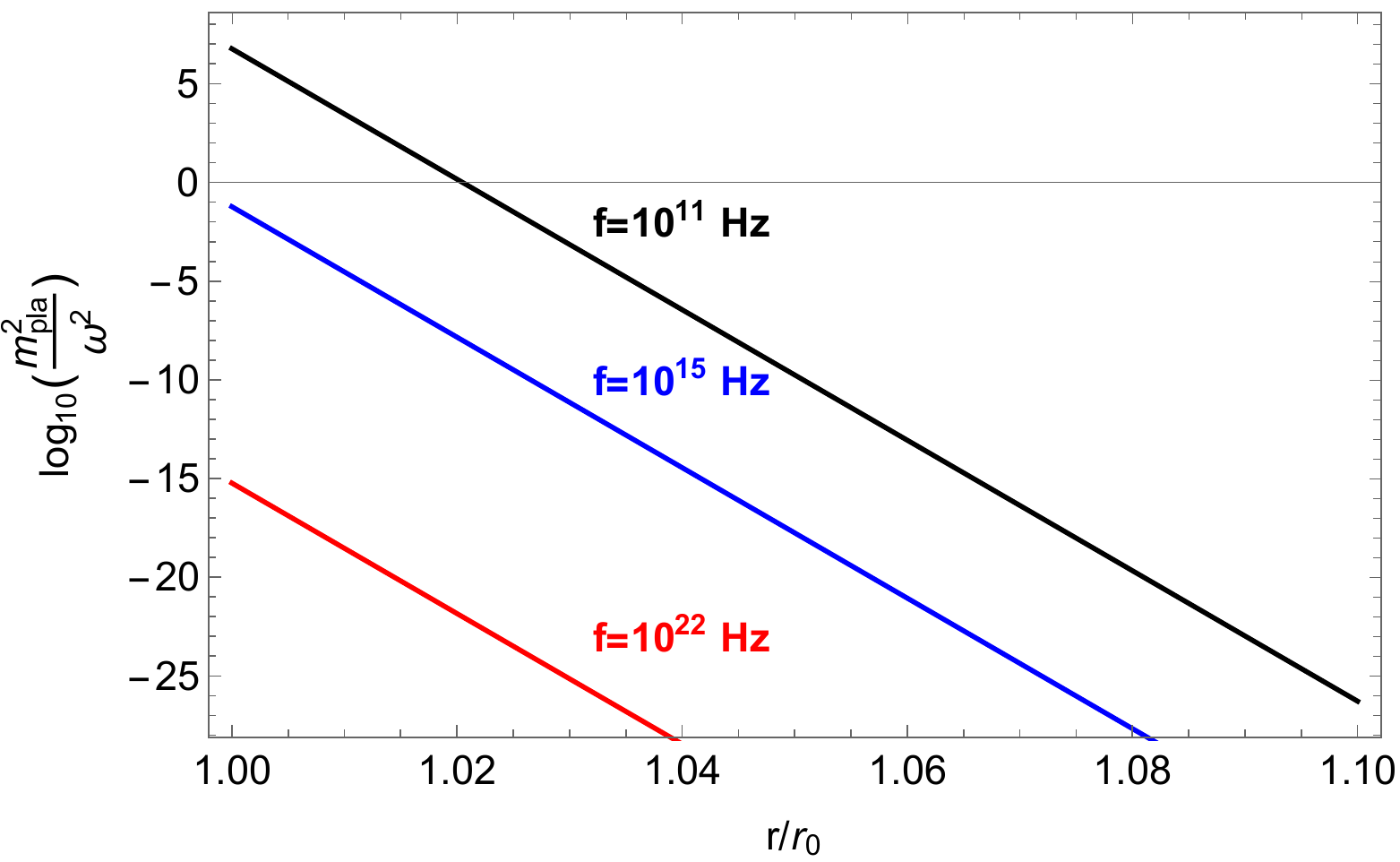}  
\caption{$m_{\rm pla}^2/\omega^2$ as a function of $r/r_0$ in the Earth's magnetosphere, along the radial direction considered in Fig.~\ref{fig:p_all}. }
   \label{wpw}
\end{figure}

For the formalism developed so far, we have neglected the plasma effect in the dielectric tensor of media,  which has been carefully analyzed in the context of axion-photon conversion in the  highly magnetized anisotropic plasma of NSs~\cite{Millar:2021gzs}. The dielectric tensor is adjusted by terms proportional to $m_{\rm pla}^2/(\gamma^3 \omega^2)$, where $\gamma$ is Lorentz factor. These terms can mix the transverse and longitudinal modes of GW-converted photons, and significantly alter the flux transfer in the direction of incident GWs. 
To look into this effect in our case, we demonstrate in Fig.~\ref{wpw} the $m_{\rm pla}^2/\omega^2$ factor as a function of $r/r_0$ for different GW frequencies in the Earth's magnetosphere. $\gamma \approx 1$ has been assumed for non-relativistic plasma in the planetary magnetosphere.
For $f\gtrsim 10^{11}\,$Hz, namely the GW frequencies considered for sensitivity analysis in Fig.~\ref{fig:limit}, we find $m_{\rm pla}^2/\omega^2\ll 1$ for $r/r_0\gtrsim 1.02$, which corresponds to an altitude $H\gtrsim 130\,$km above the Earth's surface. $m_{\rm pla}^2/\omega^2 > 1$ only occurs for $r\to r_0$ and relatively low GW frequencies, where however the relative contributions to the integrated GW-photon conversion probability for the LEO detectors  tend to be suppressed (see, e.g., Fig.~\ref{fig:dpdr} below). Therefore, for the LEO detectors with $H\gtrsim 600\,$km considered in our study, the plasma effect in the dielectric tensor of media can be safely neglected.

Due to the smallness of gravitational coupling, we have   $\Delta_{\rm M}\ll \Delta_{\gamma}$ and hence $l_{\rm osc} =2/(4\Delta_{\rm M}^2+\Delta_{\gamma}^2)^{1/2} \approx 2/\Delta_{\gamma}$ in general. This implies a weak mixing between the GWs and photons. Nonetheless, $\Delta_{\rm vac}$ and $\Delta_{\rm pla}$ have opposite signs, and a maximum mixing can be realized if the two cancel well such that $\Delta_{\gamma}\ll \Delta_{\rm M}$.
In the planetary magnetosphere, such a cancellation might be possible only for a tiny range of the GW traveling path. Its contribution to the total conversion probability is typically negligible. Instead, the ``coherent conversion'', where the conversion probability is maximized for given magnetic field and GW path, tends to be realized for a longer part of the path and hence contribute more to the GW-photon conversion.

To have a taste, in the main text we first analyze the GW-photon conversion for a radial path of the planets from zero altitude and latitude to spatial infinity. Due to the difference in spinning velocity, the plasma density of planets is described by Barometric formula 
\be
n_c^{\rm Bar}=n_{c,0} \exp \left(-\frac{r-r_0}{H_{\rm cor}} \right)
\label{n_epl}  \, ,
\ee
where $H_{\rm cor} \ll r_0$ denotes the scale height and $n_{c,0}$ denotes the surface plasma density, with $H_{\rm cor}\approx 8.4\, (27) \textrm{ km}$ and $n_{c,0}\approx 10^{26}\, (10^{29})\,\rm fm^{-3}$ for the Earth~(Jupiter)~\cite{lee2015scale,EarthSpec,JupiterSpec}, while the plasma density of NSs is described by Goldreich-Julian model~\cite{Goldreich:1969sb}. As $n_c^{\rm Bar}$ is exponentially suppressed at $r-r_0\gg H_{\rm cor}$, the plasma mass contribution is mostly negligible along the path for the GW frequency of interest. Thus, we focus on the vacuum effect and assume $\Delta_{\gamma}= \Delta_{\rm vac}$. 
$P_0$ then can be solved exactly as~\cite{Dessert:2019sgw,Fortin:2018aom,Fortin:2021sst}
\be
 	P_0 \approx \frac{(\Delta_{M}(r_0) r_0)^2}{(\Delta_{\rm vac}(r_0) r_0)^{\frac{4}{5}}}
	\left| \frac{\Gamma(\frac{2}{5}) - \Gamma(\frac{2}{5}, -\frac{i}{5} \, \Delta_{\rm vac}(r_0) r_0) }{5^{\frac{3}{5}}} \right|^2 \,,
\label{eq:Pexact}
\ee
where $\Gamma(z)$ is gamma function and $\Gamma(a,z)$ is incomplete gamma function. $ \Delta_{\rm M}(r_0)$ and $\Delta_{\rm vac}(r_0)$ are defined at the planet surface where $r=r_0$.

\begin{figure}[h]
\centering
\includegraphics[width=9cm]{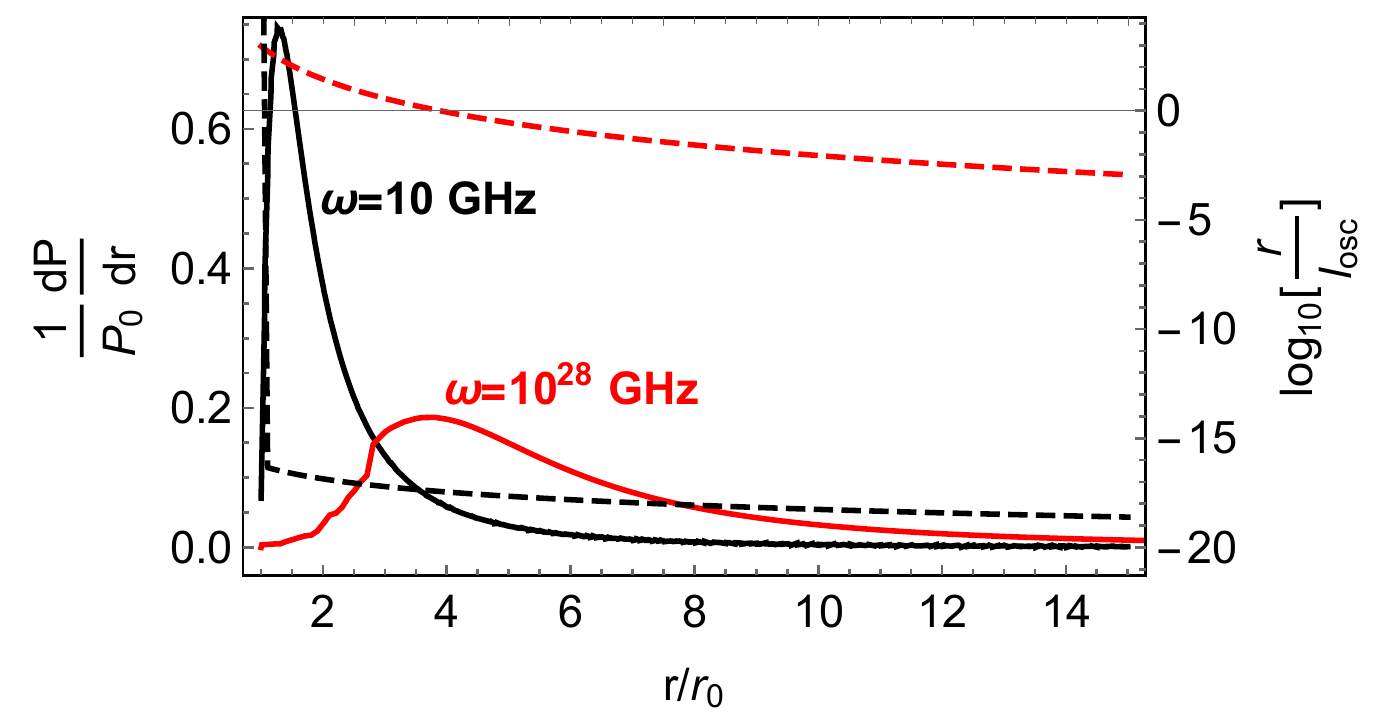}
\caption{ $\frac{1}{P_0} \frac{d P(r)}{dr}$ (left axis, solid curves) and $\text{log}_{10}[\frac{r}{l_{\rm osc}(r)}]$  (right axis, dashed curves) as a function of $r/r_0$ in the Earth’s magnetosphere ($\omega_{\textrm{tra}}\approx 10^{25}\textrm{GHz}$), along the radial direction considered in Fig.~\ref{fig:p_all}. The black and red curves denote the low-frequency ($\omega=10$ GHz) and high-frequency ($\omega=10^{28}$ GHz) scenarios, respectively.}
\label{fig:dpdr}
\end{figure}

In the low frequency limit, where $\Delta_{\rm vac}(r_0)\, r_0\lesssim 1$ and hence $r_0\lesssim l_{\rm osc}(r_0)$, the coherent conversion can be achieved near the planet surface. Eq.~\ref{eq:Pexact} then gives the first formula in Eq.~(\ref{eq:Papprox}):
\ba
P_0&\approx& \frac{1}{4}\Delta^2_{ M}(r_0) r^2_0\nonumber\\
&=& 2.1\times 10^{-32} \left(\frac{B_0}{1\,\textrm{Gauss}}\right)^2\left(\frac{r_0}{10^4\,\textrm{km}}\right)^2\, .
\label{eq:Papprox_low}
\ea
Notably, the GW-photon conversion can be coherent also in the region well above the planet surface. $\text{log}_{10}[\frac{r}{l_{\rm osc}(r)}]$ is uniformly smaller than zero for $r \gtrsim r_0$, as shown in Fig.~\ref{fig:dpdr}. But, the contributions to $P_0$ from the high-altitude magnetosphere are relatively few since the magnetic field and hence $\Delta_M(r)$ become weakened as the altitude increases. So, $P_0$ receives most of the contributions from the region near the Earth surface, yielding a sharp peak at $r \sim r_0$ for the solid black curve in Fig.~\ref{fig:dpdr}.

In the high frequency limit, where $\Delta_{\rm vac}(r_0)\, r_0\gtrsim 1$ and hence $r_0\gtrsim l_{\rm osc}(r_0)$, the coherent conversion can not be achieved near the Earth surface. Eq.~(\ref{eq:Pexact}) then gives the second formula in Eq.~(\ref{eq:Papprox}): 
\ba
P_0 &\approx&   0.7 \frac{(\Delta_{M}(r_0)\, r_0)^2}{ (\Delta_{\rm vac}(r_0) \,r_0)^{\frac{4}{5}}} 
=\frac{1}{4}\Delta^2_{M}(r_0) r^2_0 \left(\frac{\omega}{\omega_\textrm{tra}}\right)^{-\frac{4}{5}} \,,
\label{eq:Papprox_high_p} 
\ea
where the transition frequency
\begin{eqnarray}
\omega_\textrm{tra}\approx 3\times 10^6 \left(\frac{B_0}{10^{10}{\rm Gauss}}\right)^2 {\rm GHz}  
\end{eqnarray}
is determined by  
$\Delta_{\rm vac}(r_0)\, r_0\approx 1$. Eq.~(\ref{eq:Papprox_high_p}) matches to  Eq.~(\ref{eq:Papprox_low}) smoothly at $\omega=\omega_\textrm{tra}$. In this limit however the coherent conversion can be realized in the magnetosphere at high altitude, where $r=r_*>r_0$. $r_*$ can be estimated by solving $\Delta_{\rm vac}(r_*)r_*= 1$.  
With $\Delta_{M}(r)=\Delta_{M}(r_0)(r_0/r)^3$ and $\Delta_{\rm vac}(r)=\Delta_{\rm vac}(r_0)(r_0/r)^6$, we obtain $r_*= r_0(\Delta_{\rm vac}(r_0) r_0)^{1/5}$. This implies that $P_0$ here receives more contributions at $r\gtrsim r_*$. These features are shown in Fig.~\ref{fig:dpdr} for the Earth. Indeed, the contributions from the region near the Earth surface get suppressed in the high-frequency case, which leads to a relatively gentle peak for the solid red curve near $r_*/r_0\approx 10^{-5} (\omega/\text{GHz})^{1/5}$.

\begin{figure}[h]
 \centering
 \includegraphics[width=8cm]{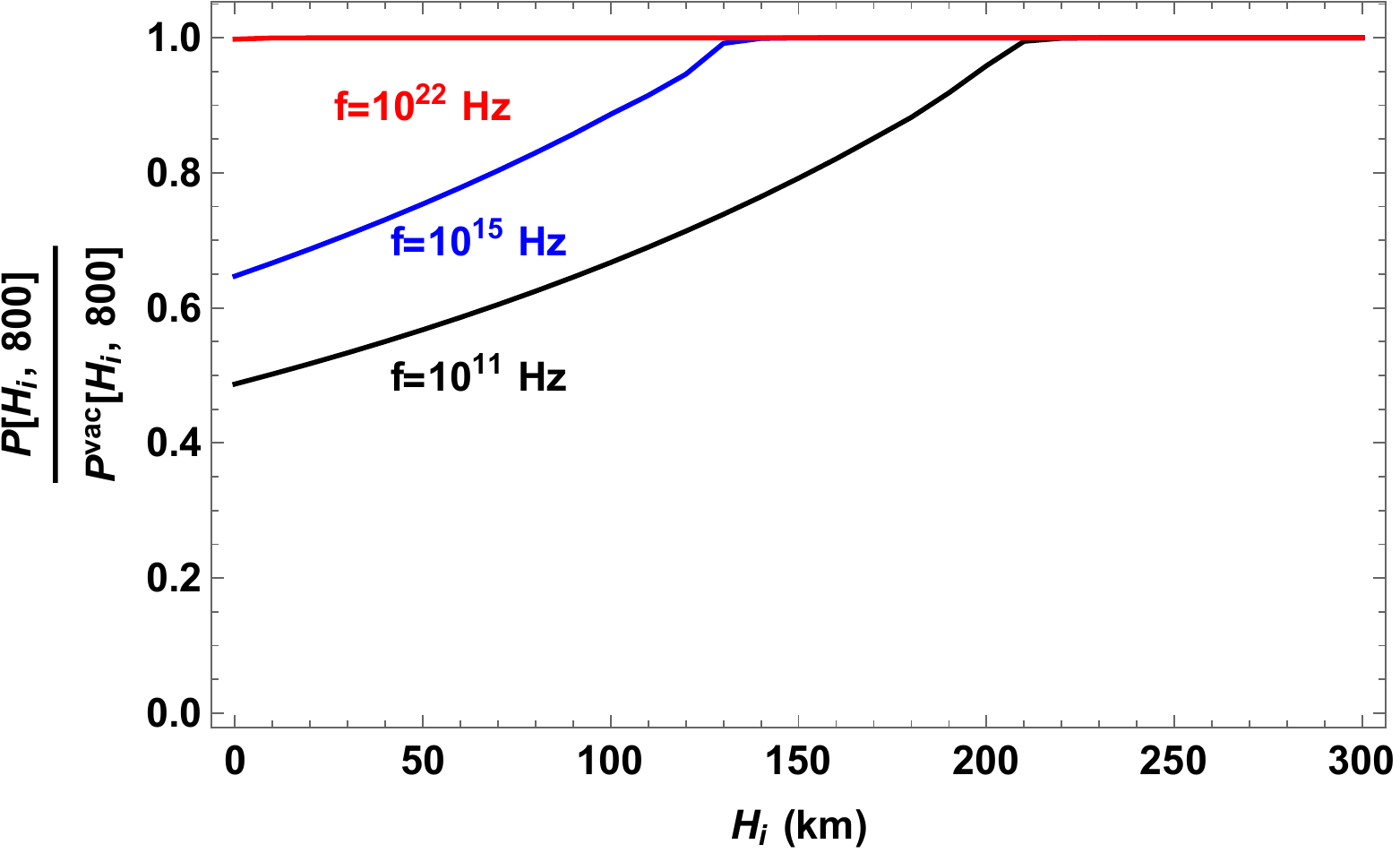}  
\caption{$\frac{P[H_i,800]}{P^{\rm vac}[H_i,800]}$ as a function of $H_i$, for a LEO detector over the equator. Here $P[H_i,800]$ and $P^{\rm vac}[H_i,800]$ are GW-photon conversion probabilities. They are calculated by integrating the radial-direction contributions from an altitude $H_i< H$ to the detector location $H=800$ km, with the plasma term $\Delta_{\rm pla}$ in $\Delta_{\gamma}$ being turned on and off respectively. 
 }
   \label{fig:ratio}
\end{figure}

The features of the Earth and Jupitor curves in Fig.~\ref{fig:p_all} can be well-explained with the formulae obtained by neglecting the plasma mass of photons in the planet magnetosphere, while these curves (together with Fig.~\ref{fig:dpdr}) are drawn based on a full calculation. Actually, we can further demonstrate that such a negligence has limited impacts only on the accuracy of results, for the GW frequency range of interest. To see this point, we demonstrate in Fig.~\ref{fig:ratio} the ratio of conversion probabilities $P[H_i,800]/P^{\rm vac}[H_i,800]$ for a typical LEO detector, as a function of $H_i$. From this figure, one can see that the $P^{\rm vac}[H_i,800]$, calculated with the plasma term $\Delta_{\rm pla}$ in $\Delta_{\gamma}$ being turned off, deviates from the full calculation $P[H_i,800]$ by a factor $\lesssim 2$ for $f>10^{11}\,$Hz. This deviation is maximized while $H_i \to 0$ and $f \to 10^{11}\,$Hz, as the plasma density gets exponentially enhanced with $r\to r_0$ and meanwhile $\Delta_{\rm pla} \propto m^2_{\rm pla}/\omega$ becomes more important for small $\omega$. Such a deviation however is too small to cause a significant impact on the sensitivity discussions in this proof-of-concept study.

\begin{figure}[h]
\centering
\includegraphics[width=8.5cm]{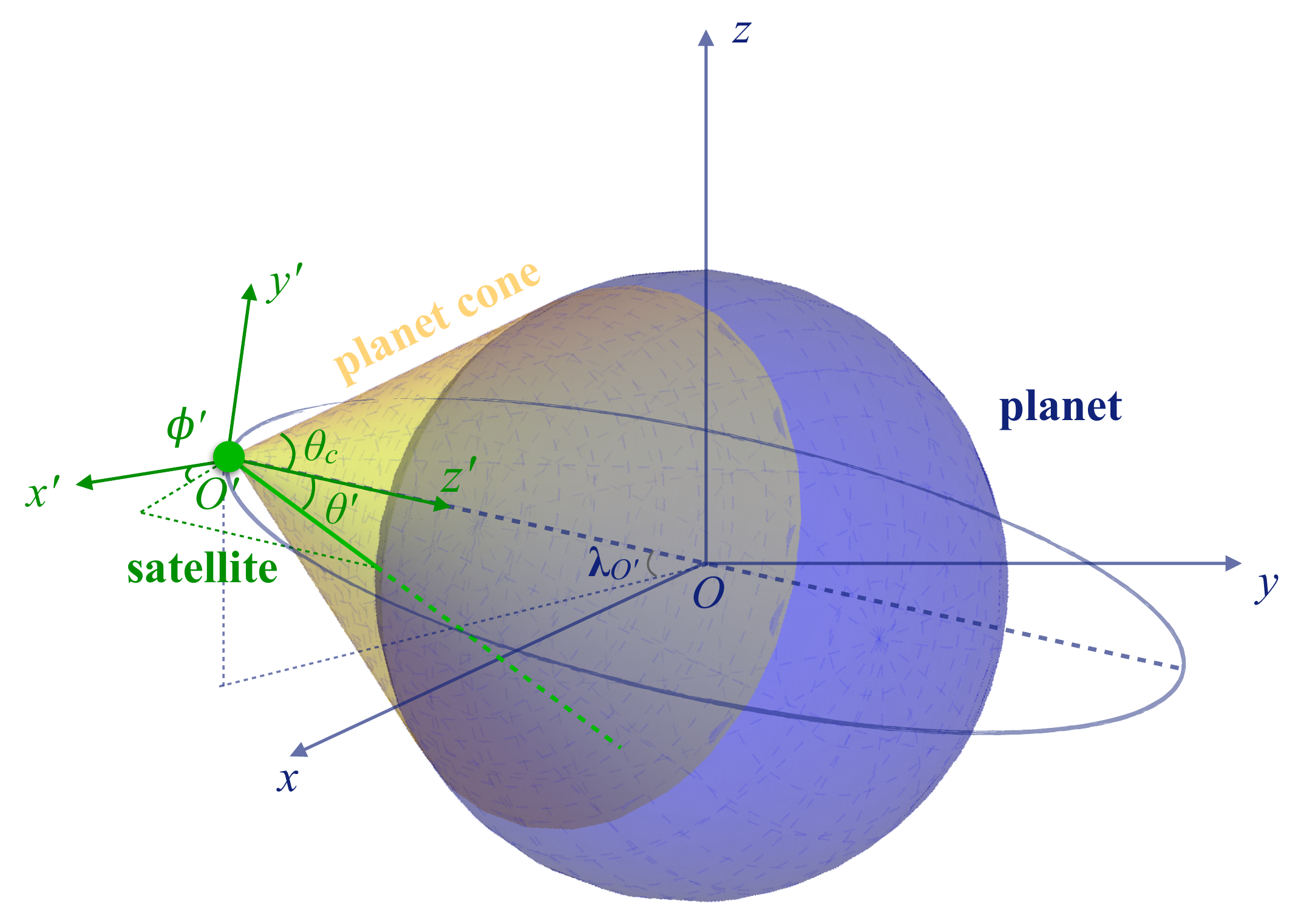}
\caption{Instant coordinate system for a satellite-based detector. $\{x,y,z\}$ and $\{x',y',z'\}$ are the coordinate systems centered at the planet center $O$ and the satellite location $O'$, respectively, with $z$ axis and $z'$ axis respectively oriented towards the planet north pole and center. $\lambda_{O'}$ denotes the latitude of the satellite. $\theta_c$ is the open angle of the PC that is tangential to the planet's surface. $\theta'$ and $\phi'$ are the polar and azimuthal angles of the line of sight (solid green line) defined in the $\{x',y',z'\}$ frame, respectively.
} 
\label{fig:coord}
\end{figure}

For sensitivity demonstration, we consider a satellite-based detector. The instant coordinate system for this detector is displayed in Fig.~\ref{fig:coord}. As the GW-photon conversion probability $P(\Omega')$ exhibits a stronger dependence on $\theta'$ than on $\phi'$, it is convenient to show its $\theta'$ dependence in Fig.~\ref{fig:p_thetaL}, by integrating out $\phi'$.

\begin{figure}[h]
\centering
\includegraphics[width=8.5cm]{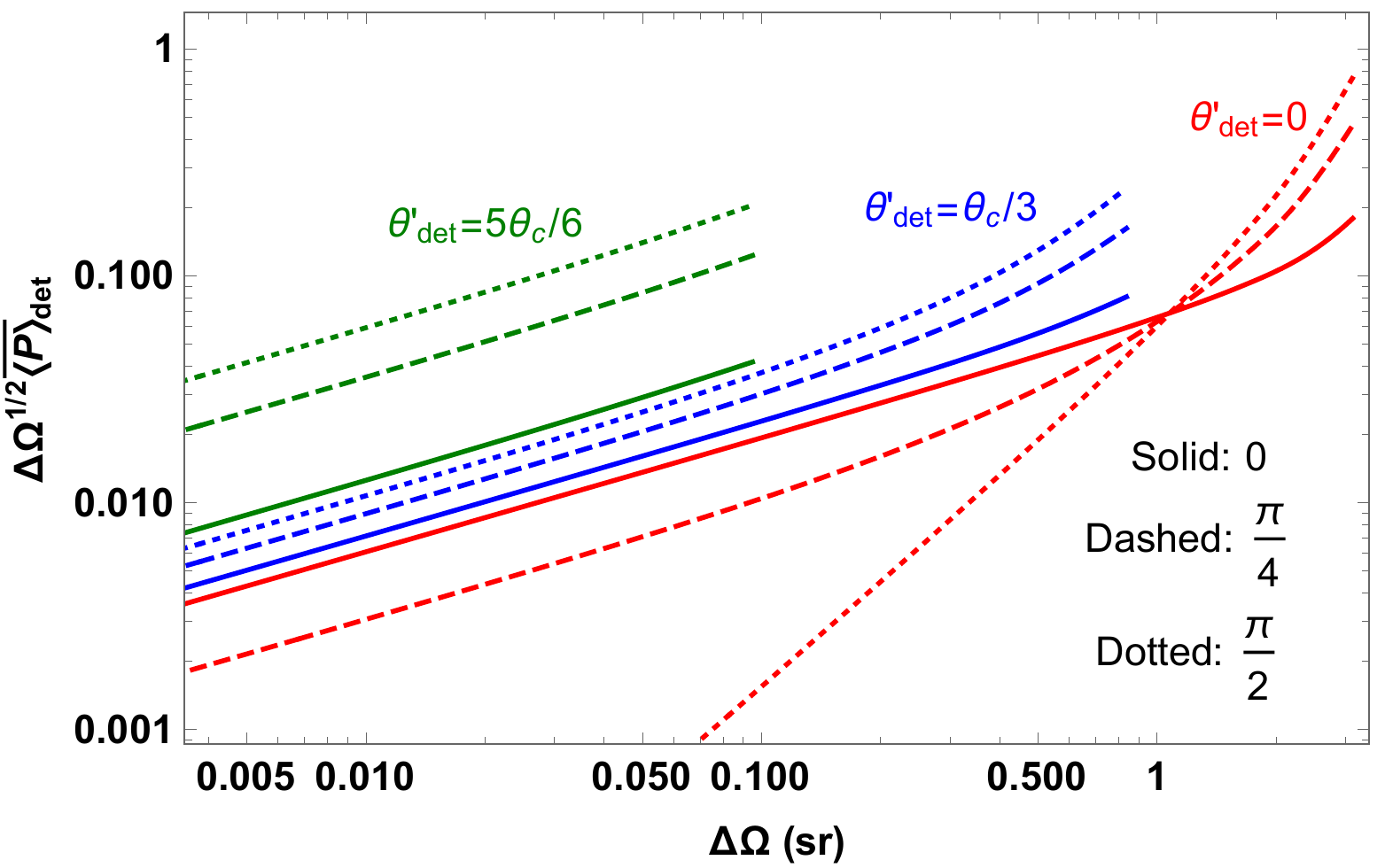}
\caption{$\Delta \Omega^{1/2} \overline{\langle P \rangle}_{\rm det}/P_0$ as a function of $\Delta \Omega$ for a satellite-based detector at $H/r_0=0.16$, where $\overline{\langle P \rangle}_{\rm det}=\int_{0}^{2\pi} {\langle P \rangle}_{\rm det}d\phi'_{\rm det}/(2\pi P_0)$ is an average $\langle P \rangle_{\rm det}$ over $\phi'_{\rm det}$. The line style and color denote the satellite latitude $\lambda_{O'}$ and the detector pointing direction $\theta'_{\rm det}$, respectively. } 
\label{fig:optOmega}
\end{figure}

 As shown in Eq.~(\ref{eq:hcUL1}), the limits on $h_c$ are sensitive to the satellite position, the detector pointing direction \{$\theta’_{\rm det}, \phi’_{\rm det}$\} and its FOV $\Delta\Omega$.
 Considering $h_{c,95\%}\propto \Delta \Omega^{1/2} \langle P \rangle_{\rm det}$, 
we show in Fig.~\ref{fig:optOmega} $\Delta \Omega^{1/2} \overline{\langle P \rangle}_{\rm det}$ as a function of $\Delta \Omega$ for different satellite latitudes and detector pointing directions $\theta'_{\rm det}$. For each curve, the quantity increases monotonically with $\Delta\Omega$ in general, and approximately scales as $\Delta\Omega^{1/2}$ for small FOV. As for the $\theta'_{\rm dec}$ dependence, the curves change more drastically for the high latitude satellite, in accordance with the angular dependence shown in Fig.~\ref{fig:p_thetaL}. $\Delta \Omega^{1/2} \overline{\langle P \rangle}_{\rm det}$ reaches its maximum when the satellite operates at  $\lambda_{O'} =\pi/2$, with the detector FOV covers the whole PC region. As the PC edge contributes more to the signal rate, a sensitivity of the same level could be also obtained for a detector with relatively small FOV but pointing to the PC edge.

\subsection{B. Low Earth Orbits for Satellites}

For a satellite-based detector, the signal and background photon fluxes are direction and time dependent. To evaluate the orbit quality accurately and hence optimize its design, one can bin the events w.r.t. direction and time while analyzing the  sensitivity reach. Eq.~(\ref{eq:DeltaSB}) is then generalized to 
 \begin{eqnarray}
 s_{ij}&=& \frac{h_c^2}{4 \pi \kappa^2} A\, \Delta t_j \Delta \omega P_{\Delta\Omega_i}(\Delta t_j)  \, , \nonumber \\
 b_{ij}&=& \phi_b (\Delta \Omega_i, \Delta t_j)  A\,\Delta t_j  \Delta \Omega_i \Delta\omega  \, , 
 \end{eqnarray}
 for the $i$-th angular and $j$-th time bin. 
Given the count of observed events $n_{ij}$, the exclusion limit on $h_c$ is set by the test statistic~\cite{Cowan:2010js}
\begin{eqnarray}\label{eq:qhc}
q(h_c)=2\sum_{i,j} \left[n_{ij}\ln \frac{n_{ij}}{ s_{ij}+ b_{ij}}+ s_{ij}+ b_{ij}-n_{ij}\right] \, ,
\end{eqnarray} 
which is Poisson-statistics-based and defined against the background + signal ($h_c$) hypothesis. 
The projected $95\%$ C.L. upper limit on $h_c$ is then obtained by requesting $q(h_{c,95\%})|_{n=b} = 2.7$, where $n_{ij}$ is assumed to be the same as the predicted background count. In the Gaussian statistics, Eq.~(\ref{eq:qhc}) is reduced to $q(h_c)=\sum_{i,j} s_{ij}^2/(b_{ij}+s_{ij})$. It can be further simplified to $q(h_c)=\sum_{i,j} s_{ij}^2/b_{ij}$ in the limit of large background, where $b_{ij}\gg s_{ij}$. This yields the criteria of $s/\sqrt{b}=1.64$  for the one-bin analysis, which has been applied for the derivation of  Eq.~(\ref{eq:hcUL1}).

\begin{table}[htbp]
\begin{center}
\resizebox{\columnwidth}{!}{%
\begin{tabular}{cccccc}
\hline\hline
     &  $H$ (km)
     &  $\theta_{\rm inc}$ 
     & $\psi$ 
     &  $T$ (mins) \\
\hline
\\
&&&&&
\\[-6.5mm]
Suzaku-like~\cite{Suzaku} \,\, & \,\, 600 \,\,  & \,\,$31.4^{\circ}$\,\, & \,\,$79.7^{\circ}$\,\,  & \,\, 90\,\,  \\  
\hline
\\
&&&&&
\\[-6.5mm]
SAFIR 2-like~\cite{Safir2}\,\, & \,\,800\,\,  & \,\,$98^{\circ}$\,\, &  \,\,$257^{\circ}$\,\,  & \,\, 101\,\,  &   \\
\hline\hline 

\end{tabular}
}
\caption{Two LEO benchmark scenarios. $H$ is orbit altitude, $\theta_{\rm inc}$ is inclination angle, $\psi$ is right ascension of the ascending node, and $T$ is orbit period. 
}
\label{tab:orbit}
\end{center}  
\end{table}\vspace{-0ex}

To have a taste on the impacts of orbit design, let us consider two benchmark scenarios defined in Tab.~\ref{tab:orbit}. 
In Scenario I, the orbit is Suzaku-like, characterized by a low inclination angle.    
As the detector is restricted to revolve in the low latitude region  ($|\lambda_{O'}|\lesssim 30^\circ$), the total time for its staying in the orbit of dark side, dubbed ``dark orbit'', varies little with the season alternation. This yields an overall fraction $f_{\rm dark}\approx 0.35$ or equivalently effective observation time $T_{\rm dark}\approx 1.1 \times 10^7\,$s for one-year revolution. 
In scenario II, the orbit is SAFIR 2-like, characterized by a high inclination angle. As the detector can reach the high-latitude region, the signal rate fluctuates more during each orbit period. The total time for its revolving in the dark orbit varies strongly with the season alternation. For one-year revolution, we have an overall fraction $f_{\rm dark}\approx 0.2$ and effective observation time $T_{\rm dark} \approx 6.3 \times 10^6\,$s.

\begin{figure}[h]
\centering
\includegraphics[width=7.5cm]{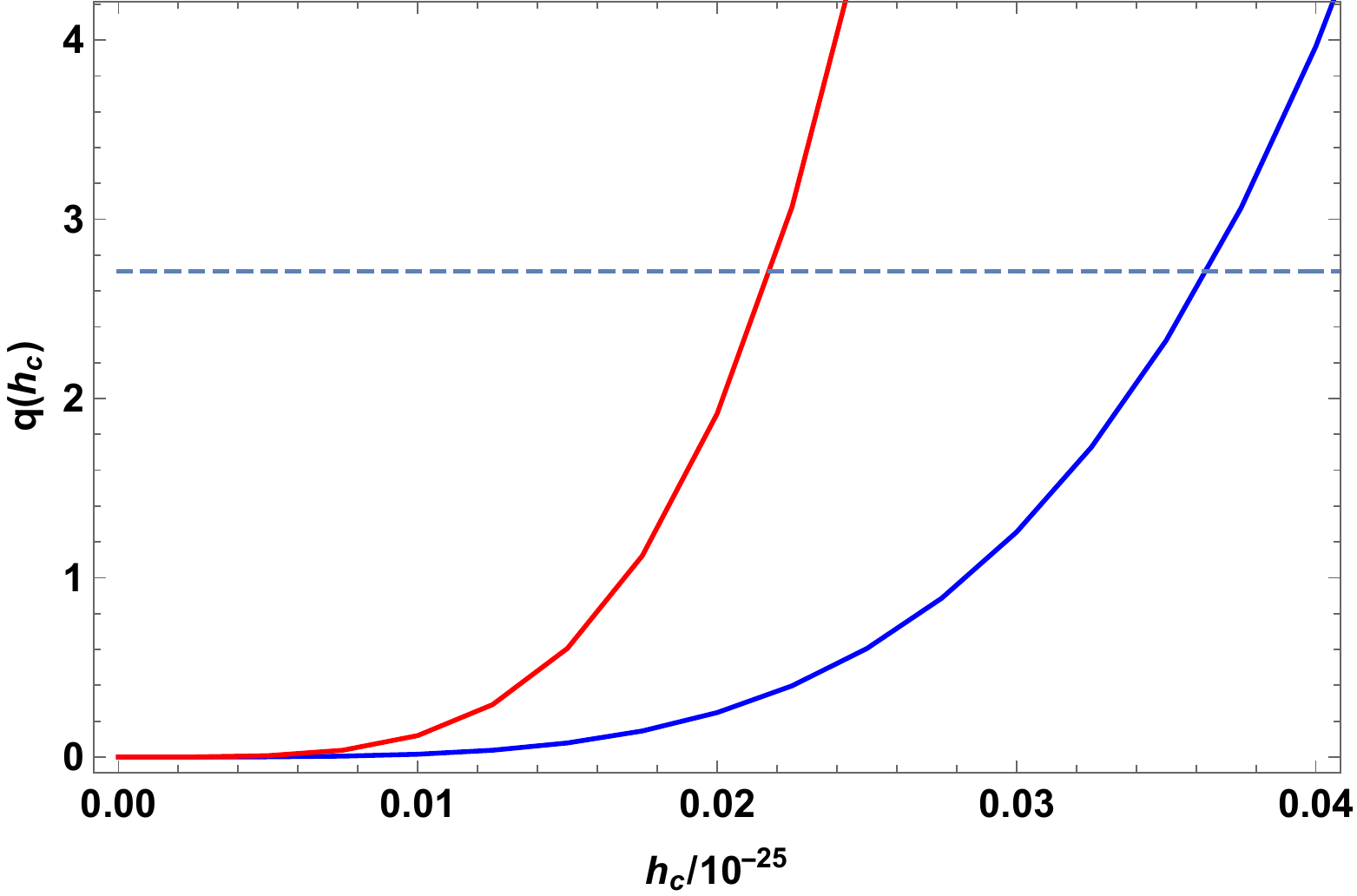}
\caption{$q(h_{c})|_{n=b}$ as a function of $h_c$ in Suzaku-like (blue) and SAFIR 2-like (red) scenarios for 10-year revolution. The dashed line denotes the $q(h_{c})|_{n=b}$ value at 95\% C.L.   
} 
\label{fig:qhc}
\end{figure}

To bin the event count in each benchmark scenario, we divide the dark orbit by a time interval $\Delta t = 120$\,s. Also, we consider the full PC region as the detector FOV and separate it into four concentric circular belts. 
We then evaluate $q(h_{c})|_{n=b}$ by taking a summation of $s_{ij}$ and $b_{ij}$ over the time-angular bins for 10-year revolution. 
Finally we display the test statistic $q(h_{c})|_{n=b}$ in Fig.~\ref{fig:qhc} as a function of $h_c$ for these two scenarios (with $A=250\,\rm cm^2$). The SAFIR 2-like orbit has a better sensitivity performance, although it spends less time in the dark orbit. This could be attributed to a larger GW-photon conversion probability off the PC center while the detector revolves in the high-latitude region (see Fig.~\ref{fig:p_thetaL}). Also, as one merit of high-inclination orbit, the dark-orbit time is mainly contributed by the polar-night season, enabling a more efficient usage of the telescope resources.

\subsection{C. Recast Limits from Laboratory Experiments}
\label{app:labexp}

The limits obtained in the laboratory experiments for axion detection can be recast to constrain the HFGWs (for a review, see~\cite{Aggarwal:2020olq}). Since the magnetic field in these experiments is often confined inside a straight-long cylinder pipe, the conversion probability of stochastic HFGWs into photons exhibits a strong angular dependence. This will necessarily influence the estimate of the GW-converted photon flux in Eq.~(\ref{eq:fluxg}). 

\begin{figure}[h]
\centering
\includegraphics[width=8.5cm]{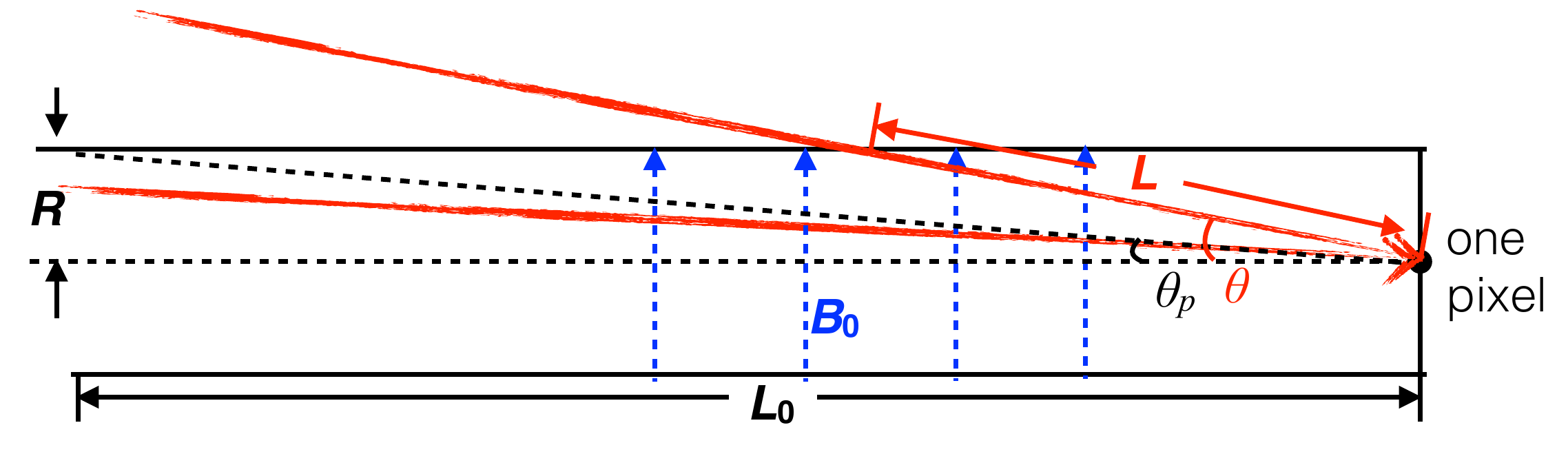}
\caption{Schematic for the GW-photon conversion in a straight-long cylinder pipe which has a length $L_0$, a  cross-sectional area $A=\pi R^2$ and an internal magnetic field ${\bf B}_0$ (with $B_0=|{\bf B}_0|$). The black dot denotes the central pixel of the detector at the right end of the pipe. The red lines denote the incoming GWs. These GWs travel in the pipe by a distance $L$, in a direction with $0 \leq \theta < \pi/2$. $\theta_p\approx R/L_0\ll 1$ denotes the opening angle of pipe. 
} 
\label{fig:pipe}
\end{figure}

Explicitly, let us consider the coherent conversion in an experimental facility shown in Fig.~\ref{fig:pipe}. For the central pixel of its detector, the angular integral in Eq.~(\ref{eq:fluxg}) reduces to an integration over $\theta$, which can be put into two parts. The first one accounts for the signals received in the opening angle, where $0\leq \theta \leq \theta_p$. As
$P(\theta)\propto  L^2 B_t^2=(L_0/\cos\theta)^2(B_0\cos\theta)^2=L_0^2 B_0^2 $, the conversion probability is insensitive to $\theta$. 
However, for the signals outside the opening angle, where $\theta_p < \theta < \pi/2$, $P(\theta)$ shows a strong dependence on $\theta$, as $P(\theta)\propto L^2 B_t^2=(R/\sin\theta)^2(B_0\cos\theta)^2=R^2 B_0^2/\tan^2\theta$. The integrated flux at this pixel is then  
\begin{eqnarray}
\label{eq:Phig_pipe}
\Phi_\gamma &\approx& \frac{\kappa^2 B_0^2}{4} \frac{\Delta \omega\, h_c^2}{4\pi\kappa^2}\int_{0}^{2\pi} d\phi \Big(L_0^2\int_{0}^{\theta_p} \sin\theta d\theta  \nonumber \\&+& R^2\int_{\theta_p}^{\pi/2}\frac{\sin\theta d\theta}{\tan^2\theta}  \Big) \nonumber \\
&\approx&\frac{1}{8}\Delta \omega h_c^2 B_0^2L_0^2 \theta_p^2 \ln \frac{1}{\theta_p}\, .
\end{eqnarray}
A limit of $\theta_p\ll 1$ is taken in the last line. 
Note, the $\phi$ dependence of $B_t$ has not been considered in this derivation. It can be shown that this effect is tiny for  Eq.~(\ref{eq:Phig_pipe}) in the small $\theta_p$ limit. Similar derivation can be applied to other pixels of the detector. 
For the experiments listed in Tab.~\ref{tab:labexp}, where $\theta_p$ is consistently small, we have verified numerically that the results for all pixels can be approximated by the leading-order term in Eq.~(\ref{eq:Phig_pipe}), with a deviation  up to $\mathcal{O}(10\%)$. Thus, the GW-converted photon flux in the detector is estimated as
\be\label{eq:compare}
 \Phi_\gamma \sim \Phi^0_\gamma \frac{\Delta\Omega_p}{\pi}\frac{1}{4}\ln\frac{\pi}{\Delta\Omega_p}\,,
 \ee
where for a narrow frequency band
\begin{eqnarray}
 \Phi^0_\gamma
\approx \frac{1}{4}{B_0^2\, L_0^2\, h_c^2\, \Delta\omega} \, . 
\label{Flux}
\end{eqnarray}
Note that $\Phi_\gamma$ is subject to a suppression of pipe geometry denoted by  
\begin{eqnarray}
\Delta\Omega_p \approx \frac{A}{L_0^2} \approx \pi \theta_p^2 \, .
\end{eqnarray}

\begin{table}[h]
\begin{center}
\begin{tabular}{ccccccc}
\hline\hline

     &  $N_{\rm exp}$ (mHz)
     &  $A$ (${\rm m}^2$)
     &  $L_0$ (m) 
     & $B_0$ (T) 
     & $\Delta f$ (Hz)\\
\hline
\\
&&&&&&
\\[-6.5mm]

CAST\,\, & \,\, 0.15 \,\, & \,\, 0.0029\,\, & \,\,$9.3$\,\, & \,\,$9$\,\,  & $10^{18}$ \\

\hline
\\
&&&&&&
\\[-6.5mm]

ALPS I\,\, & \,\, 0.61 \,\, & \,\, 0.005\,\, & \,\,$9$\,\, & \,\,$5$\,\,  & $9\times 10^{14}$ \\

\hline
\\
&&&&&&
\\[-6.5mm]

OSQAR II\,\, & \,\, 1.1 \,\, & \,\, 0.005\,\, & \,\,$14$\,\, & \,\,$9$\,\,  & $10^{15}$ \\

\hline\hline

\end{tabular}
\caption{Baseline parameters for some laboratory experiments~\cite{Ejlli:2019bqj} relevant to this study.}
\label{tab:labexp}
\end{center}  \
\end{table}\vspace{-0ex}

The upper limits on $h_c$ can be obtained by comparing the signal rate $\Delta s/\Delta t$ with the observed event rate $N_{\rm exp}$  (see Tab.~\ref{tab:labexp}), which gives 
\begin{eqnarray}\label{eq:hcpipe}
h_{c,95\%}\approx \frac{4}{B_0 A} \sqrt{\frac{\pi N_{\rm exp}}{\Delta\omega\ln(\pi L_0^2/A)}}\,.
\end{eqnarray}
Then we have the limits shown in Fig.~\ref{fig:limit}: 
$h_{c,95\%}\sim 8\times 10^{-26}$ for the CAST (see Eq.~(\ref{eq:hcCAST}) also), and $h_{c,95\%}\sim 5\times 10^{-23}\,$ and $4\times 10^{-23}\,$ for the ALPS I and the OSQAR~II, respectively.
These recast limits are less powerful than the estimates in Ref.~\cite{Ejlli:2019bqj}. This discrepancy can be attributed to the difference of signal photon flux $\Phi_\gamma$ in Eq.~(\ref{eq:compare}) from the one obtained in Ref.~\cite{Ejlli:2019bqj}.
As noticed, the factor $\sim \Delta\Omega_p/(4\pi)$ appearing in Eq.~(\ref{eq:compare}), representing a suppression by the small opening angle of long pipe, is missing in Eq.~(2) of  Ref.~\cite{Ejlli:2019bqj}, where the energy flux of signal photons  $\Phi_\gamma^{\rm graph}$ is approximately $\sim \omega \Phi_\gamma^0$. This discrepancy highlights the impacts of the angular dependence of GW-photon conversion probability and the range of detector FOV, as depicted in Fig.~\ref{fig:pipe}.


\end{document}